\documentclass[12pt,headings=big,numbers=noenddot,DIV=14,a4paper]{scrartcl}%

\pdfoutput=1

\usepackage{amsmath,amssymb,amsfonts}
\usepackage[normal,font=small,labelfont=bf,labelsep=period]{caption}
\usepackage[pdftex]{color,graphicx}
\usepackage{subfigure}
\usepackage[english]{babel}
\usepackage[compress]{cite}
\addtokomafont{disposition}{\rmfamily\boldmath}

\usepackage[dvipsnames]{xcolor}
\definecolor{darkblue}{rgb}{0,0.2,0.6}
\definecolor{darkgreen}{rgb}{0,0.4,0}

\usepackage[linktoc=page,bookmarks=false,colorlinks=false,
linkbordercolor=RoyalBlue,citebordercolor=ForestGreen,urlbordercolor=CornflowerBlue]{hyperref}

\numberwithin{equation}{section}
\DeclareFontFamily{OT1}{pzc}{}
\DeclareFontShape{OT1}{pzc}{m}{it}{<-> s * [1.10] pzcmi7t}{}
\DeclareMathAlphabet{\mathpzc}{OT1}{pzc}{m}{it}
\DeclareMathOperator{\tr}{tr} % Trace

\title{\begin{flushright}
       \mbox{\normalsize\rm CERN-PH-TH-2013-170}
       \end{flushright}
       \vskip 30pt
       \textcolor{black}{One or more Higgs bosons?}}
\author{\normalsize Riccardo
Barbieri$^a$, Dario Buttazzo$^{a,b}$, Kristjan Kannike$^{a,c}$, Filippo Sala$^{a,d}$, Andrea Tesi$^a$}
\date{\normalsize{\it  $^a$Scuola Normale Superiore and INFN, Piazza dei Cavalieri 7, 56126 Pisa, Italy\\
$^b$CERN Theory Division, CH-1211 Geneva 23, Switzerland\\
$^c$National Institute of Chemical Physics and Biophysics, R\"avala 10, Tallinn 10143, Estonia\\
$^d$Theoretical Physics Group, Lawrence Berkeley National Laboratory, Berkeley, CA 94720}}

\begin{document}
\begin{titlepage}

\maketitle
\thispagestyle{empty}

\begin{abstract}
\centerline{\bf Abstract}\medskip
\noindent
Now that one has been found, the search for signs of more scalars is a primary task of  current and future experiments. In the motivated hypothesis that the extra Higgs bosons of the next-to-minimal supersymmetric Standard Model (NMSSM) be the lightest new particles around, we outline a possible overall strategy to search for signs of the $CP$-even states. This work complements Ref. \cite{Barbieri:2013hxa}.
\end{abstract}

\vfill
\noindent\line(1,0){188}\\\medskip
\footnotesize{E-mail: \scriptsize\tt{\href{mailto:barbieri@sns.it}{barbieri@sns.it}, \href{mailto:dario.buttazzo@sns.it}{dario.buttazzo@sns.it}, \href{mailto:kannike@cern.ch}{kannike@cern.ch}, \href{mailto:filippo.sala@sns.it}{filippo.sala@sns.it}, \href{mailto:andrea.tesi@sns.it}{andrea.tesi@sns.it}}}

\end{titlepage}

\section{Introduction}
\label{sec1}

The discovery of the/a Higgs boson, $h_{\text{LHC}}$, with a mass of about 126 GeV and Standard-Model-like properties  \cite{Aad:2012tfa,Chatrchyan:2012ufa,ATLAS,CMS,tevatron:2013} raises a clear question: Is it the coronation of the Standard Model (SM) or a first step into yet largely unexplored territory? The answer to this question, whose relation with the absence so far of any signal of new physics does not need to be illustrated, is in some sense paradoxical. While  the  newly found resonance  completes the SM spectrum, thus representing a major milestone in the entire history of particle physics, there are still good reasons to think that its discovery may not be the end of the story. The many well-known problems that the finding of the resonance, viewed as the Higgs boson of the SM,  leaves unresolved are one order of reasons.
Quite independently and in fact on more general grounds, now that a scalar particle has been found, one may wonder if and why it should be alone and not part of an extended Higgs system. Since we know of no strong argument in favour of a single scalar particle, it is justified to think that the search for  signs of extra scalars is a primary task of  current and future experiments.

A motivated example of an extended Higgs system is the one of the next-to-minimal supersymmetric Standard Model (NMSSM), which adds to a usual Higgs doublet one further doublet and one complex singlet under the $\mathrm{SU}(2)\times \mathrm{U}(1)$ gauge group, all parts of corresponding chiral super-multiplets, so that to allow a supersymmetric gauge-invariant Yukawa-like coupling $\lambda S H_u H_d$ \cite{Fayet:1974pd} (see \cite{Ellwanger:2009dp} for a review and references). In spite of the presence of (broken) supersymmetry, the Higgs sector of the NMSSM is not free from the problem common to the introduction of a  scalar sector in any gauge theory:
the significant number of free parameters that describe their masses and interactions, of which the Yukawa couplings in the SM are a prototype example. In turn this explains the difficulty of a simple enough description of the related phenomenology as well as the extended literature on the subject.\footnote{For a partial list of recent references, see \cite{Hall:2011aa,Ellwanger:2011aa,Cao:2012fz,Jeong:2012ma,Agashe:2012zq-1,Belanger:2012tt,Choi:2012he,King:2012tr,DAgnolo:2012mj,Gupta:2012fy,Gherghetta:2012gb,Kang:2013rj,Cheung:2013bn,Cheng:2013fma,Badziak:2013bda,Bhattacherjee:2013vga,Ellwanger:2013ova}.}
The purpose of this work is to outline a possible overall strategy  to search for signs of the $CP$-even extra-states of the NMSSM Higgs sector. This paper must be viewed as a complement of \cite{Barbieri:2013hxa}, to which we add: i) the consideration of the case in which one state exists below $h_{\text{LHC}}$; ii) the expected sensitivity on the  overall parameter space of the measurements of the signal strengths of $h_{\text{LHC}}$ at LHC14 with their projected errors; iii) the consideration of the impact of the EWPT on the different situations. To keep things  comprehensive  we will have to make some simplifying assumptions, that we shall be 
careful to specify whenever needed.

\section{Reference equations}
\label{sec2}

For ease of the reader we summarize in this Section the definitions and the reference equations that we shall use to describe the relation between the physical observables and the parameters of a generic NMSSM.

Assuming  a negligibly small $CP$-violation in the Higgs sector, the original scalar fields $\mathcal{H} = (H_d^0, H_u^0, S)^T$ are related to the   three $CP$-even physical mass eigenstates $ \mathcal{H}_{\rm ph} = (h_3, h_1, h_2)^T$ by
\begin{equation}
\mathcal{H} = R^{12}_{\alpha} R^{23}_{\gamma} R^{13}_{\sigma} \mathcal{H}_{\rm ph} \equiv R \mathcal{H}_{\rm ph},
\label{rotation_matrix}
\end{equation}
where $R^{ij}_\theta$ is the rotation matrix in the $i j$ sector by the angle $\theta = \alpha,\gamma,\sigma$. We shall identify the resonance found at LHC with $h_1$.

In full generality the mixing angles $\delta\equiv\alpha-\beta +\pi/2, \gamma, \sigma$ can be expressed in terms of the masses $m_{h_1,h_2,h_3}$ and $m_{H^{\pm}}$, the charged Higgs boson mass, as\cite{Barbieri:2013hxa}\footnote{Notice that Eq.~\eqref{eq:sin:2alpha:general} is completely equivalent to the expression for $\sin 2\alpha$ in Eq. (2.10) of Ref.~\cite{Barbieri:2013hxa}.}

\begin{align}
  s_\gamma^{2} &=  \frac{ \det M^{2} + m_{h_1}^{2} (m_{h_1}^{2} - \tr M^{2})}{(m_{h_1}^{2} - m_{h_2}^{2}) 
  (m_{h_1}^{2} - m_{h_3}^{2}) }, 
  \label{eq:sin:gamma:general}
  \\
  s_\sigma^{2} &= \frac{m_{h_2}^{2} - m_{h_1}^{2}}{m_{h_2}^{2} - m_{h_3}^{2}} \; \frac{ \det M^{2} + m_{h_3}^{2} (m_{h_3}^{2} - \tr M^{2}) }
  { \det M^{2} - m_{h_2}^{2} m_{h_3}^{2} + m_{h_1}^{2} (m_{h_2}^{2} + m_{h_3}^{2} - \tr M^{2}) },
  \label{eq:sin:sigma:general}
  \\
     s_{2\delta}&=
  \Big[ 
    2 s_\sigma c_\sigma s_\gamma \left(m_{h_3}^2-m_{h_2}^2\right) \left(2 \tilde M^2_{11}-m_{h_1}^2c_\gamma^2 -m_{h_2}^2(s_\gamma^2+s_\sigma^2c_\gamma^2) - m_{h_3}^2(c_\sigma^2+s_\gamma^2 s_\sigma^2)\right) \notag
    \\ 
    & +2 \tilde M^2_{12} \left(m_{h_3}^2 \left(c_\sigma^2-s_\gamma^2 s_\sigma^2\right)+m_{h_2}^2 \left(s_\sigma^2-s_\gamma^2  c_\sigma^2\right)-m_{h_1}^2 c_\gamma^2 \right)
  \Big]
  \notag \\
  & \times \Big[ \left(m_{h_3}^2-m_{h_2}^2 s_\gamma^2- m_{h_1}^2 c_\gamma^2\right)^2
  +\left(m_{h_2}^2-m_{h_3}^2\right)^2 c_\gamma^4 s_\sigma^4 \notag\\
   &+2 \left(m_{h_2}^2-m_{h_3}^2\right) \left(m_{h_3}^2+m_{h_2}^2 s_\gamma^2-m_{h_1}^2 \left(1+s_\gamma^2\right)\right) c_\gamma^2 s_\sigma^2
  \Big]^{-1}, \label{eq:sin:2alpha:general}
   \end{align}  
where   $s_\theta = \sin{\theta}, c_\theta = \cos{\theta}$, $M^2$ is the $2\times 2$ submatrix in the $1 2$ sector of the full $3\times 3$ squared mass matrix ${\cal M}^2$ of the neutral $CP$-even scalars in the ${\cal H}$ basis
\begin{equation}
M^2 = \left(
\begin{array}{ccc}
m_Z^2 c^2_\beta+m_A^2 s^2_\beta & \left(2 v^2 \lambda ^2-m_A^2-m_Z^2\right) c_\beta s_\beta  \\
 \left(2 v^2 \lambda ^2-m_A^2-m_Z^2\right) c_\beta s_\beta & m_A^2 c^2_\beta+m_Z^2 s^2_\beta +\delta_t^2 
\end{array}
\right)
\label{2x2_matrix}
\end{equation}
and $\tilde M^2= R_{\beta-\pi/2}M^2 R_{\beta-\pi/2}^t$ in Eq.~\eqref{eq:sin:2alpha:general}. In Eq.~\eqref{2x2_matrix} 
\begin{equation}
\label{mHcharged}
m_A^2 = m_{H^{\pm}}^2 - m_W^2 +\lambda^2 v^2,
\end{equation}
where  $v \simeq 174$ GeV, and
\begin{equation}
\delta_t^2 \equiv \Delta_t^2/ s^2_\beta
\label{delta-t}
\end{equation}
is the well-known effect of the top-stop loop corrections to the quartic coupling of $H_u$. We neglect the analogous correction to Eq. \eqref{mHcharged}, which lowers $m_{H^{\pm}}$ by less than 3 GeV for stop masses below 1 TeV. More importantly we have also not included in Eq. \eqref{2x2_matrix} the one loop corrections to the $12$ and $11$ entries, respectively proportional to the first and second power of $(\mu A_t)/\langle m_{\tilde{t}}^2\rangle$, to which we shall return.

We shall in particular be interested in two limiting cases:
\begin{itemize}
\item $H$ decoupled: $m_{h_3} \gg m_{h_1,h_2}$ and $\sigma, \delta \equiv \alpha - \beta +\pi/2 \rightarrow 0$,

\item Singlet decoupled:  $m_{h_2} \gg m_{h_1,h_3}$ and $\sigma, \gamma \rightarrow 0$,
\end{itemize}
but we  use Eqs. (\ref{eq:sin:gamma:general}, \ref{eq:sin:sigma:general}, \ref{eq:sin:2alpha:general}) to control the size of the deviations from the limiting cases when the heavier mass is lowered. In the two respective cases the reference equations are
\begin{itemize}
\item $H$ decoupled:
\begin{equation}
s^2_{\gamma}= \frac{m_{hh}^2-m_{h_1}^2}{m_{h_2}^2-m_{h_1}^2},
\label{sin2gamma}
\end{equation}
where 
\begin{equation}\label{mhh}
m_{hh}^2 = m_Z^2 c_{2\beta}^2 + \lambda^2 v^2 s_{2\beta}^2 + \Delta_t^2;
\end{equation}

\item Singlet decoupled: 
\begin{align}
\label{sin2alpha}
s_{2\alpha} &= s_{ 2\beta} \; \frac{2\lambda^2 v^2-m_Z^2-m_A^2|_{m_{h_1}}}{m_A^2|_{m_{h_1}} +m_Z^2 +\delta_t^2 -2m_{h_1}^2},\\
m_{h_3}^2&= m_A^2|_{m_{h_1}}+m_Z^2 +\delta_t^2 -m_{h_1}^2,
\label{mh3}
\end{align}
where
\begin{equation}\label{mA_mh}
m_A^2\big|_{m_{h_1}}=\frac{\lambda^2v^2(\lambda^2v^2-m_Z^2)s^2_{2\beta}-m_{h_1}^2(m_{h_1}^2-m_Z^2-\delta_t^2)-m_Z^2\delta_t^2 c^2_\beta}{m_{hh}^2-m_{h_1}^2}.
\end{equation}
\end{itemize}
All the equations in this section are valid in a generic NMSSM. Specific versions of it may limit the range of the physical parameters $m_{h_{1,2,3}}, m_{H^\pm}$ and $\alpha, \gamma, \sigma$ but cannot affect any of these equations.

\section{Singlet decoupled}
\label{sec3}
\begin{figure}[t!]
\begin{center}
\includegraphics[width=0.48\textwidth]{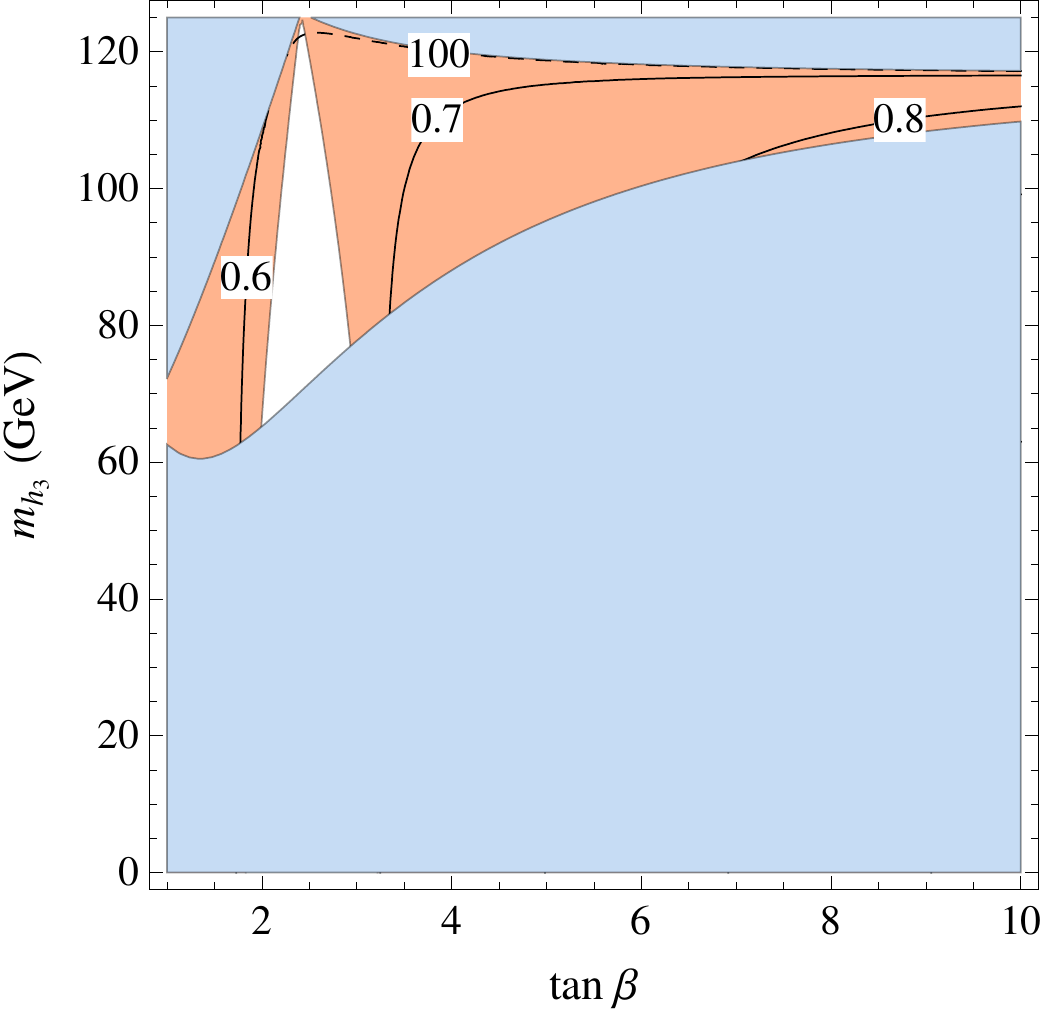}\hfill
\includegraphics[width=0.48\textwidth]{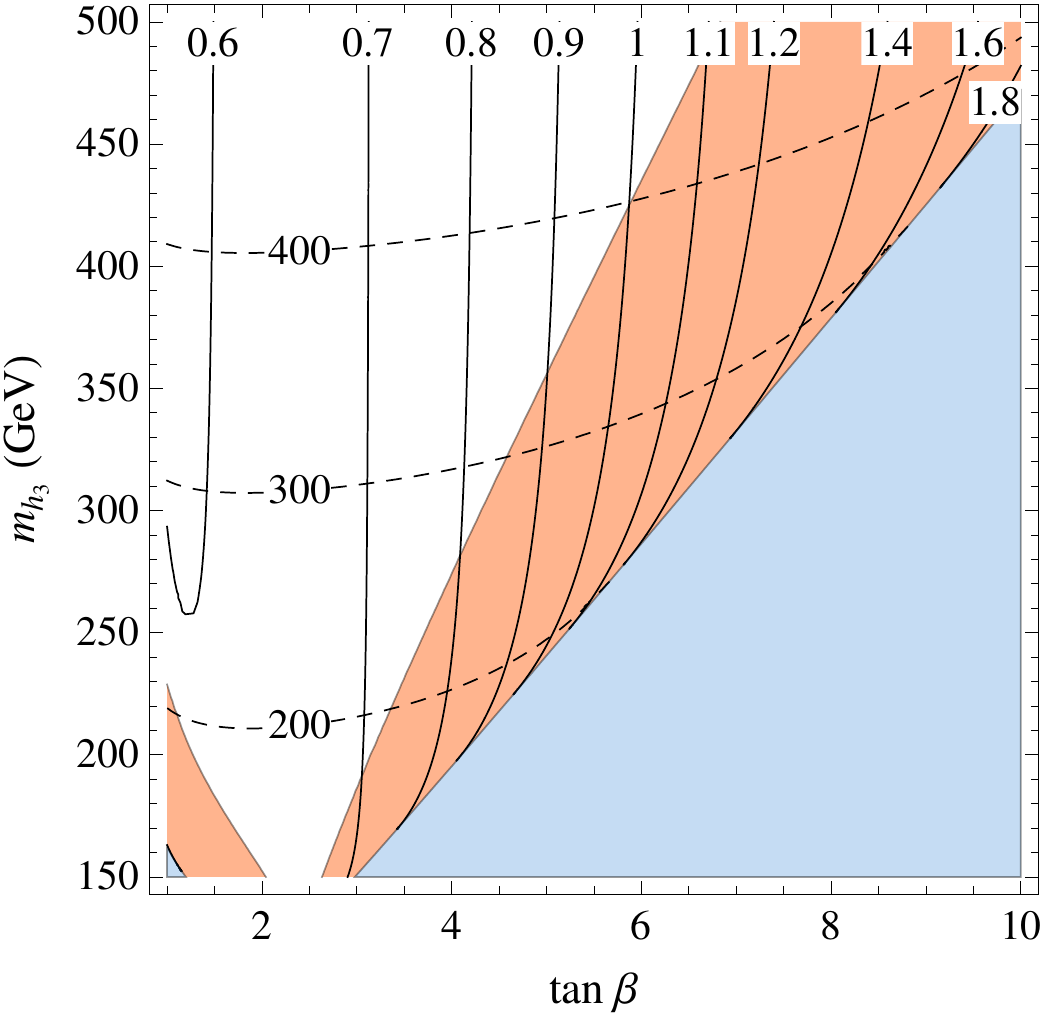}
\caption{\label{fig1} Singlet decoupled. Isolines of $\lambda$ (solid) and $m_{H^\pm}$ (dashed). Left: $h_{\rm LHC}>h_3$. Right: $h_{\rm LHC}<h_3$. The orange region is excluded at 95\%C.L. by the experimental data for the signal strengths of $h_1 = h_{\rm LHC}$. The blue region is unphysical.}
\end{center}
\end{figure}
From Eqs.~\eqref{sin2alpha}-\eqref{mA_mh} and \eqref{mHcharged}, since $m_{h_1}$ is known, $m_{h_3}, m_{H^+} $ and the angle $\delta$ are functions of  $(\tan{\beta}, \lambda, \Delta_t)$. From our point of view the main motivation for considering the NMSSM is in the possibility to account for the mass of $h_{\text{LHC}}$ with not too big values of the stop masses. For this reason  we take $ \Delta_t = 75$ GeV, which can be obtained, e.g., for an average stop mass of about 700 GeV. In turn, as it will be seen momentarily, the consistency of Eqs.  \eqref{sin2alpha}-\eqref{mA_mh} requires not too small values of the coupling $\lambda$.
It turns out in fact that for any value of $ \Delta_t \lesssim 85$ GeV,  the dependence on $\Delta_t$ itself can be neglected, so that $m_{h_3}, m_{H^\pm} $ and  $\delta$ are determined by $\tan{\beta}$ and $\lambda$ only.  For the same reason it is legitimate to neglect the one loop corrections to the $11$ and $12$ entries of the mass matrix, Eq. \eqref{2x2_matrix}, as long as $(\mu A_
t)/\langle m_{\tilde{t}}^2\rangle \lesssim 1$, which is again motivated by naturalness.

From all this we can represent in Fig. \ref{fig1} the allowed regions in the plane $(\tan{\beta}, m_{h_3})$ and the isolines of $\lambda$ and $m_{H^\pm}$ both for $h_3 < h_{\text{LHC}} (< h_3(=S))$ and for $h_{\text{LHC}} < h_3 (< h_3(=S))$, already considered in Ref. \cite{Barbieri:2013hxa}. At the same time the knowledge of $\delta$ in every point of the same $(\tan{\beta}, m_{h_3})$ plane fixes the couplings of $h_3$ and  $ h_{\text{LHC}}$, which allows to draw the currently excluded regions from the measurements of the signal strengths of $h_{\text{LHC}}$. 
We do not include any supersymmetric loop effect other than the ones that give rise to Eq.~\eqref{2x2_matrix}.
As in Ref. \cite{Barbieri:2013hxa}, to make the fit of all the data collected so far from ATLAS, CMS and Tevatron, we adapt the code provided by the authors of Ref. \cite{Giardino:2013bma}. Negative searches at LHC of $h_3 \rightarrow \bar{\tau} \tau$ may also exclude a further portion of the 
parameter space for $h_3 > h_{\text{LHC}}$. Note, as anticipated, that in every case $\lambda$ is bound to be above about $0.6$. To go to lower values of $\lambda$ would require considering $\Delta_t \gtrsim  85$ GeV, i.e. heavier stops. On the other hand in this singlet-decoupled case lowering $\lambda$ and raising $\Delta_t $ makes the NMSSM close to the minimal supersymmetric Standard Model (MSSM), to which we shall return.

When drawing  the currently excluded regions in Fig. \ref{fig1}, we are not considering the possible decays of $h_{\text{LHC}}$ and/or of $h_3$ into invisible particles, such as dark matter,  or into any undetected final state, because of background, like, e.g., a pair of light pseudo-scalars. The existence of such decays, however,  would not alter in any significant way the excluded regions from the measurements of the signal strengths of $h_{\text{LHC}}$, which would all be modified by a common factor $(1 + \Gamma_{\rm inv}/\Gamma_{\rm vis})^{-1}$. This is because the inclusion in the fit of the LHC  data of an invisible branching ratio of $h_{\text{LHC}}$, $\mathrm{BR}_{\rm inv}$, leaves essentially unchanged the allowed range for $\delta$ at different $\tan{\beta}$ values, provided $\mathrm{BR}_{\rm inv} \lesssim 0.2$.
\begin{figure}[t]
\begin{center}
\includegraphics[width=0.48\textwidth]{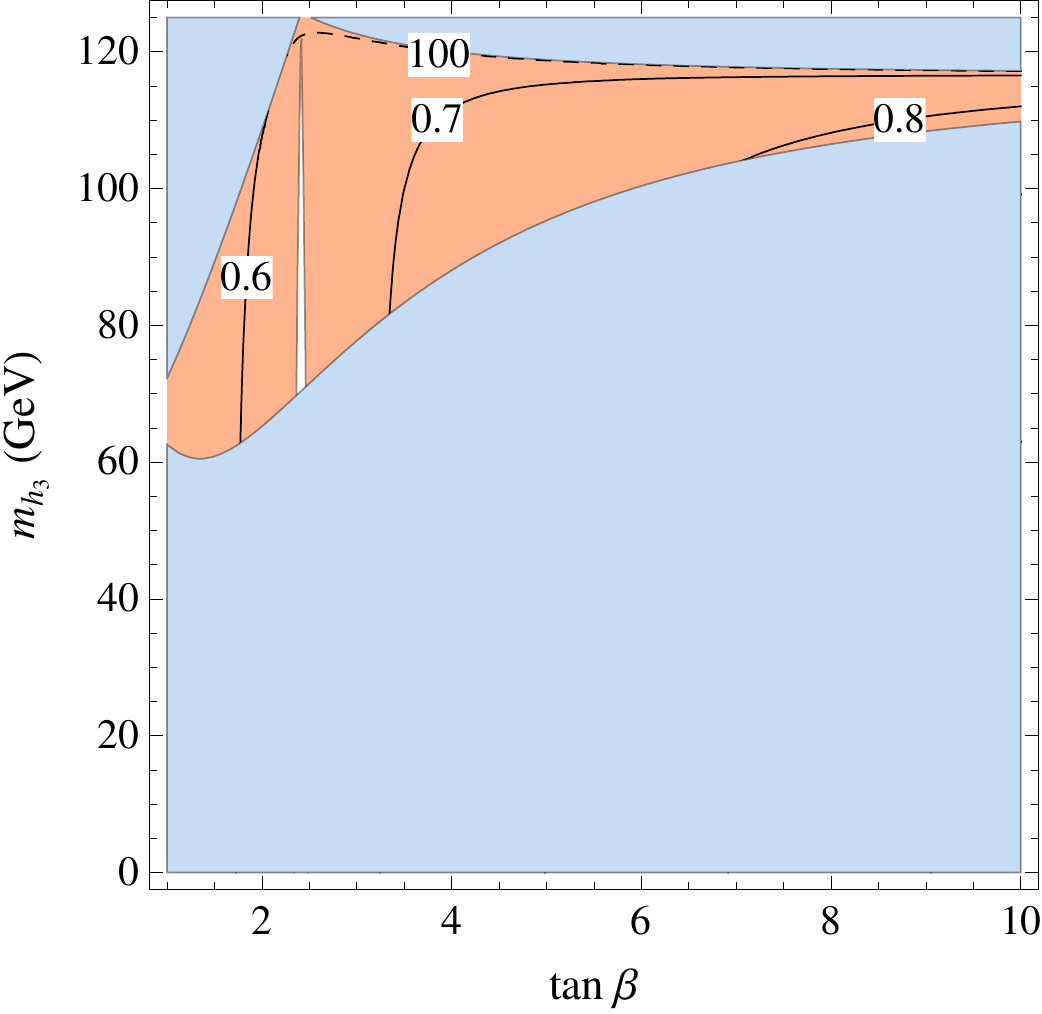}\hfill
\includegraphics[width=0.48\textwidth]{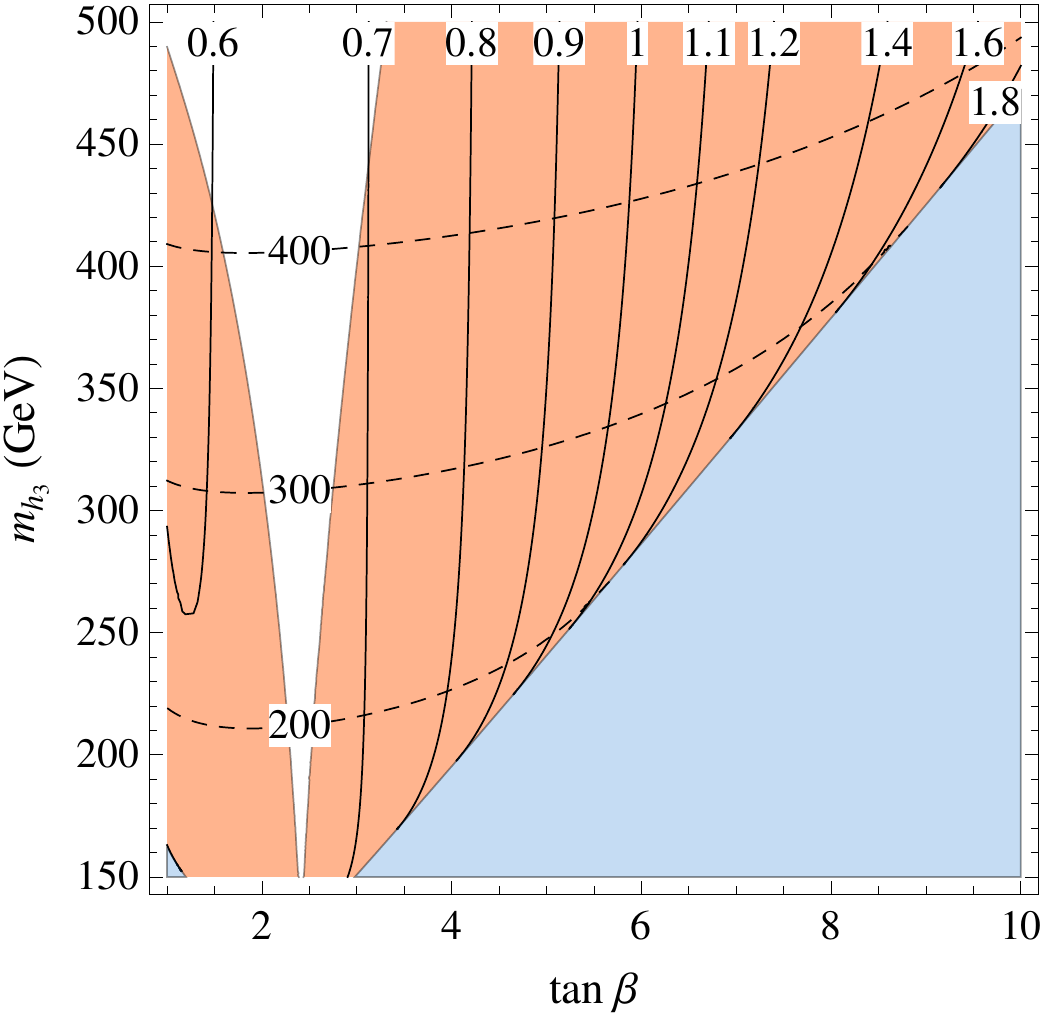}
\caption{\label{fig2} Singlet decoupled. Isolines of $\lambda$ (solid) and $m_{H^\pm}$ (dashed). Left: $h_{\rm LHC}>h_3$. Right: $h_{\rm LHC}<h_3$. The orange region would be excluded at 95\%C.L. by the experimental data for the signal strengths of $h_1 = h_{\rm LHC}$ with SM central values and projected errors at the LHC14 as discussed in the text. The blue region is unphysical.}
\end{center}
\end{figure}

The significant constraint set on Fig. \ref{fig1} by the  current measurements of the signal strengths of $h_{\text{LHC}}$ suggests that an improvement of such measurements, as foreseen in the coming stage of LHC, could lead to an effective exploration of most of the relevant parameter space. 
To quantify this we have considered the impact on the fit of the measurements of the signal strengths of $h_{\text{LHC}}$ with the projected errors at LHC14 with $300~\mathrm{fb}^{-1}$ by ATLAS\cite{ATLAS-collaboration:2012iza} and CMS\cite{CMS-14}, shown in Table \ref{tab1}.
The result  is shown in Fig. \ref{fig2}, again both for $h_3 < h_{\text{LHC}} (< h_2(= S))$ and for $h_{\text{LHC}} < h_3 (< h_2(= S))$, assuming SM central values for the signal strengths.

\begin{table}[h]
\begin{center}
\begin{tabular}{ccc}
& ATLAS & CMS \\
\hline
$h \to \gamma \gamma$ & 0.16 & 0.15 \\
$h \to Z Z$ & 0.15 & 0.11 \\
$h \to W W$ & 0.30 & 0.14 \\
$V h \to V b \bar{b}$ & -- & 0.17 \\
$h \to \tau \tau$ & 0.24 & 0.11 \\
$h \to \mu \mu$ & 0.52 & -- \\
\end{tabular}
\caption{\label{tab1}Projected uncertainties of the measurements of the signal strengths of $h_{\text{LHC}}$, normalized to the SM, at the 14 TeV LHC with $300~\mathrm{fb}^{-1}$.}
\end{center}
\end{table}
\begin{figure}[t]
\begin{center}
\includegraphics[width=0.48\textwidth]{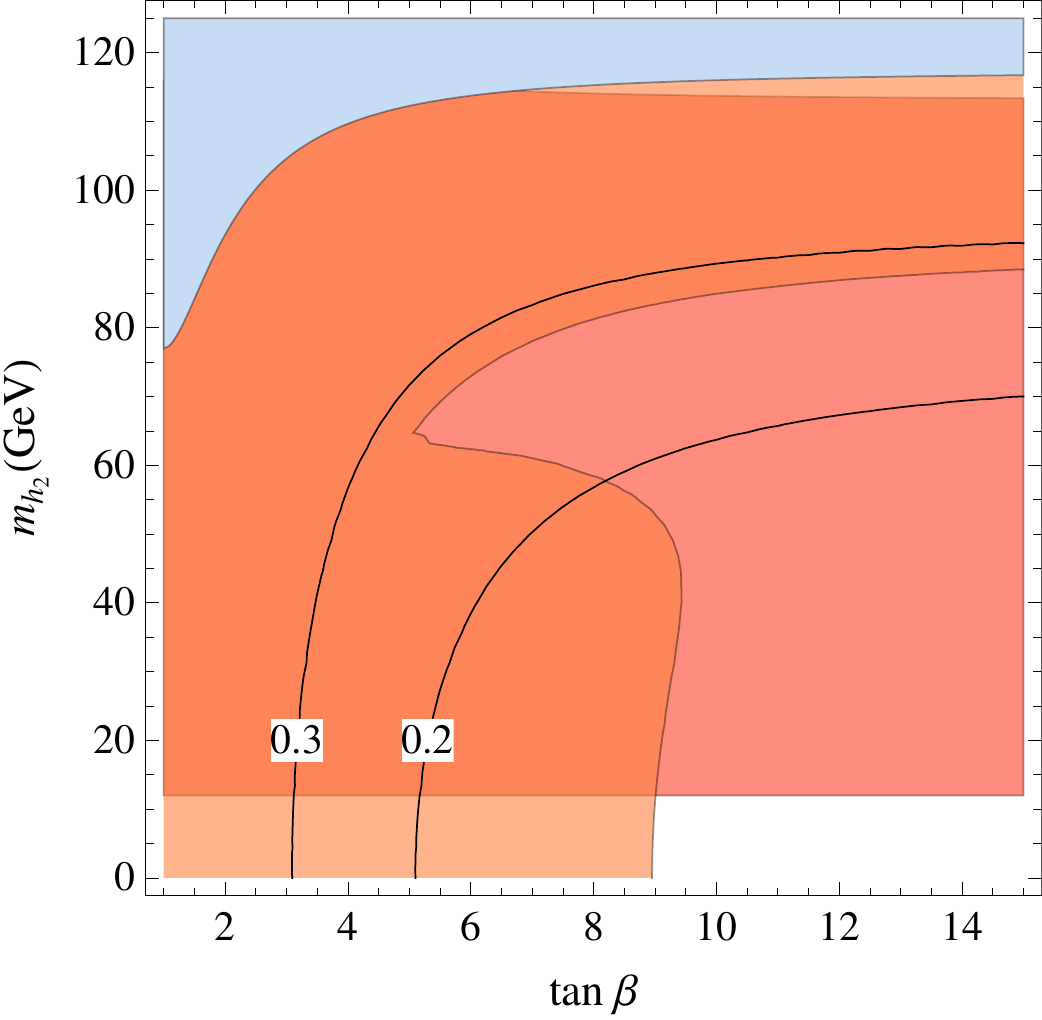}\hfill
\includegraphics[width=0.48\textwidth]{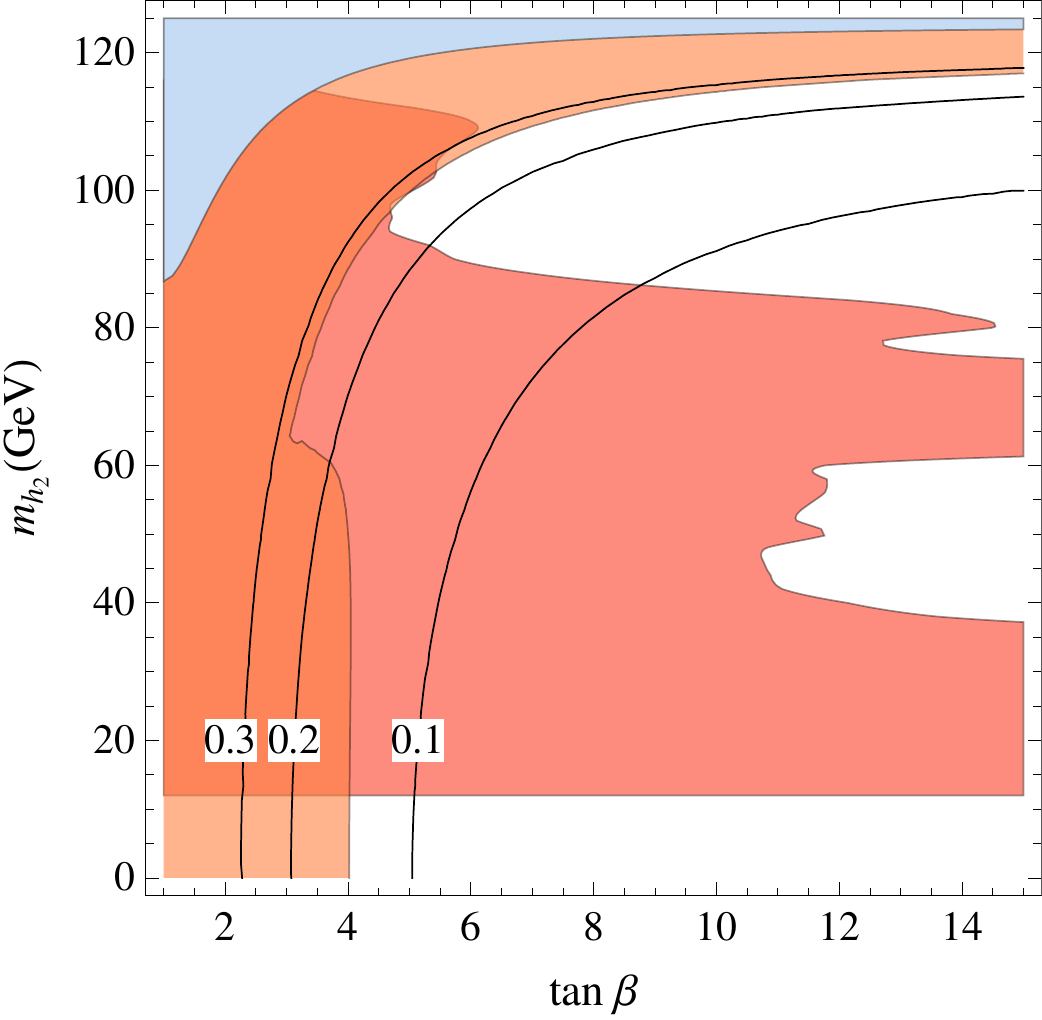}
\caption{\label{fig3} $H$-decoupled. Isolines of $s^2_\gamma$. $\lambda=0.1$ and $v_S=v$. Left: $\Delta_t=75$ GeV. Right: $\Delta_t=85$ GeV. Orange and blue regions as in Fig.~\ref{fig1}. The red region is excluded by LEP direct searches for $h_2\to b\bar b$.}
\end{center}
\end{figure}
Needless to say, the direct search of the extra $CP$-even states will be essential either in presence of a possible indirect evidence from the signal strengths or to fully cover the parameter space for $h_3 > h_{\text{LHC}}$. To this end, under the stated assumptions, all production cross sections and branching ratios for the $h_3$ state are determined in every point of the $(\tan{\beta}, m_{h_3})$ plane.

\section{$H$-decoupled}
\label{sec4}
\begin{figure}[t]
\begin{center}
\includegraphics[width=0.48\textwidth]{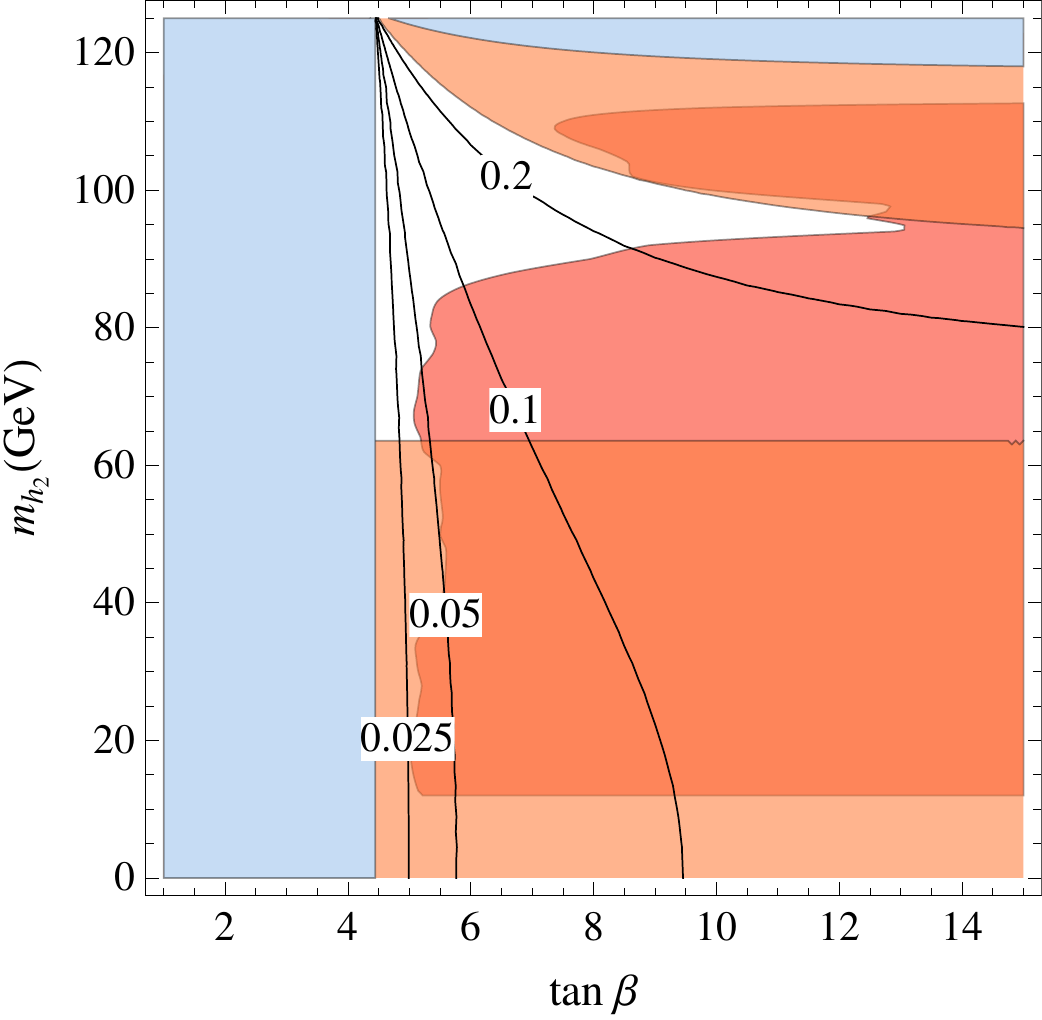}\hfill
\includegraphics[width=0.48\textwidth]{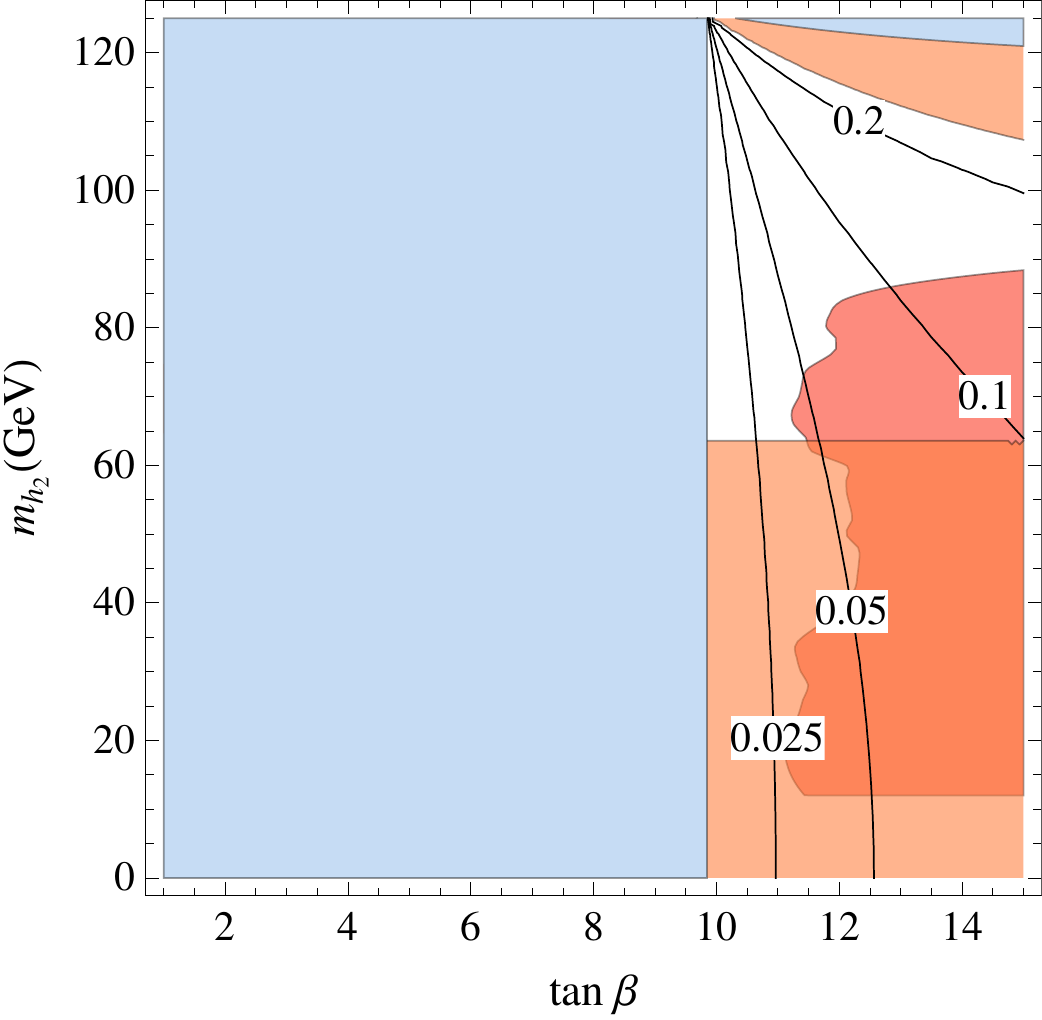}
\caption{\label{fig4} $H$-decoupled. Isolines of $s^2_\gamma$. $\Delta_t=75$ GeV and $v_S=v$. Left: $\lambda=0.8$. Right: $\lambda=1.4$. Orange and blue regions as in Fig.~\ref{fig1}. The red region is excluded by LEP direct searches for $h_2\to b\bar b$.}
\end{center}
\end{figure}
As we are going to see, the situation changes significantly when considering the $H$-decoupled case where the singlet $S$ mixes with the doublet with SM couplings. By comparing Eqs. \eqref{sin2gamma} with Eqs.\eqref{sin2alpha}, note first that in this case there is only a single relation between the mixing angle $\gamma$ and the mass of the extra $CP$-even state $m_{h_2}$, involving $\tan{\beta}, \lambda$ and
$\Delta_t$. Since the case $h_{\text{LHC}} < h_2 (< h_3(= H))$ has been extensively discussed in Ref. \cite{Barbieri:2013hxa}, here we concentrate on the case of $h_2 < h_{\text{LHC}} (< h_3(= H))$ and we consider both the low and the large $\lambda$ case.

The low $\lambda$ case $(\lambda = 0.1)$ is shown in Fig. \ref{fig3} for two values of $\Delta_t$ together with the isolines of $s^2_{\gamma}$.
Due to the singlet nature of $S$ it is straightforward to see that  the couplings of $h_1 = h_{\text{LHC}}$ and $h_2$ to fermions or to vector-boson pairs, $VV = WW, ZZ$, normalized to the same couplings of the SM Higgs boson, are given by
\begin{equation}
\frac{g_{h_1ff}}{g^{\text{SM}}_{hff}} = \frac{g_{h_1VV}}{g^{\text{SM}}_{hVV}}= c_\gamma, ~~~~~
\frac{g_{h_2ff}}{g^{\text{SM}}_{hff}}= \frac{g_{h_2VV}}{g^{\text{SM}}_{hVV}}= - s_\gamma.
\end{equation}
As a consequence for $m_{h_2} > m_{h_{\text{LHC}}}/2$
none of the branching ratios of  $h_1 = h_{\text{LHC}}$ and $h_2$ get modified with respect to the  ones of the SM Higgs boson with the corresponding mass, whereas their production cross sections are reduced by a common factor $c_\gamma^2$ or $s^2_\gamma$ respectively  for  $h_1 = h_{\text{LHC}}$ and $h_2$. The current fit of the signal strengths measured at LHC constrain $s^2_\gamma < 0.22$ at $95\%$ C.L., which explains the lighter excluded regions in Fig. \ref{fig3}. The red regions are due to the negative searches of $h_2\rightarrow \bar{b} b$ at LEP \cite{Schael:2006cr}.  As in the previous case we do not include any invisible decay mode except for $h_{\text{LHC}}\rightarrow h_2 h_2$ when kinematically allowed.\footnote{To include $h_{\text{LHC}}\rightarrow h_2 h_2$ we rely on the triple Higgs couplings as computed by retaining only the $\lambda^2$-contributions.
This is a defendable approximation for $\lambda$ close to unity, where $h_{\text{LHC}}\rightarrow h_2 h_2$ is important. In the low $\lambda$ case the $\lambda^2$-approximation can only be taken as indicative, but there $h_{\text{LHC}}\rightarrow h_2 h_2$ is less important.} Here an invisible branching ratio  of $h_{\text{LHC}}$, $\mathrm{BR}_{\rm inv}$, would strengthen the bound on the mixing angle to $s^2_\gamma < (0.22 - 0.78 \mathrm{BR}_{\rm inv})$.

For $\lambda$  close to unity we take as in the singlet-decoupled case  $\Delta_t = 75$ GeV, but any choice  lower than this would not change the conclusions. The currently allowed region is shown in Fig.~\ref{fig4} for two values of $\lambda$. Note that, for large $\lambda$, no solution is possible at low enough $\tan{\beta}$, since, before mixing, $m^2_{hh}$ in Eq.~\eqref{mhh} has to be below the mass squared of $h_{\text{LHC}}$.

\begin{figure}[t]
\begin{center}
\includegraphics[width=0.48\textwidth]{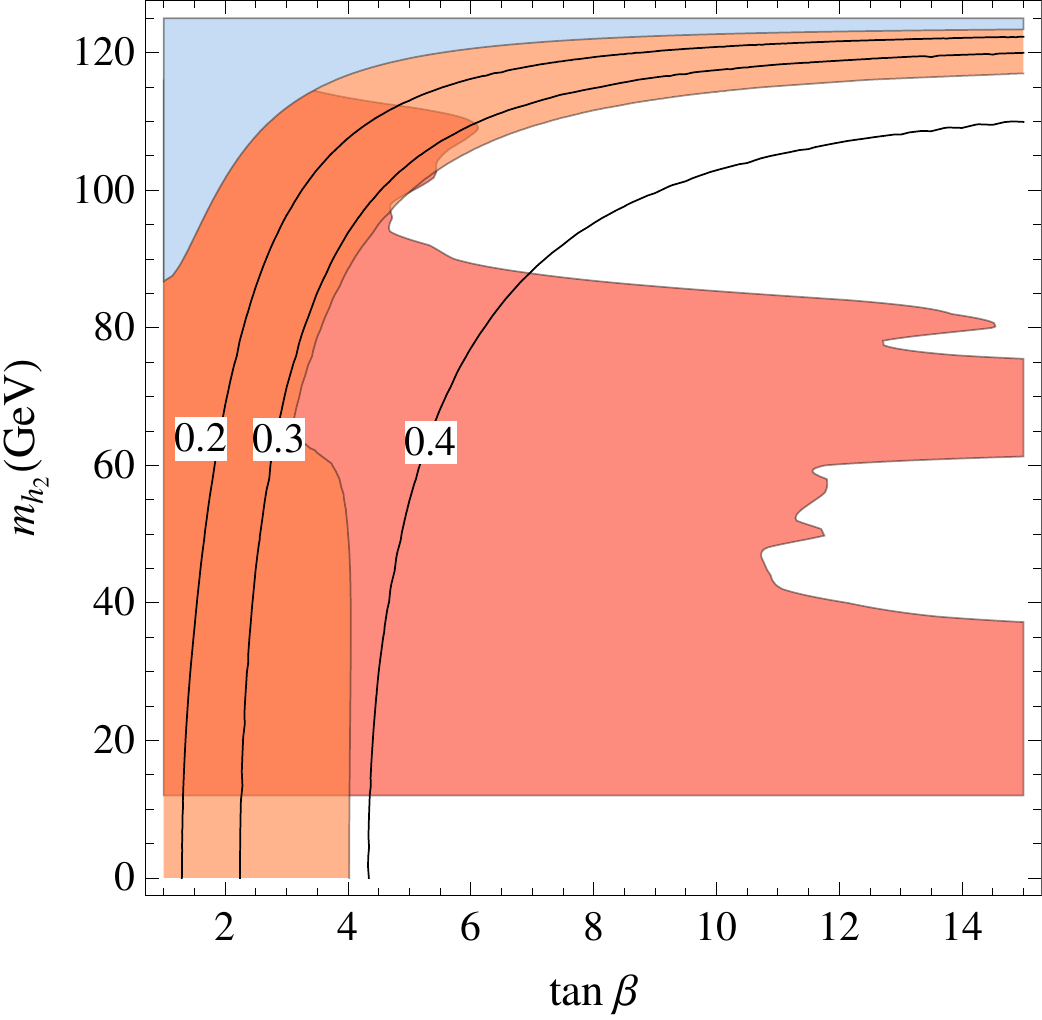}\hfill
\includegraphics[width=0.48\textwidth]{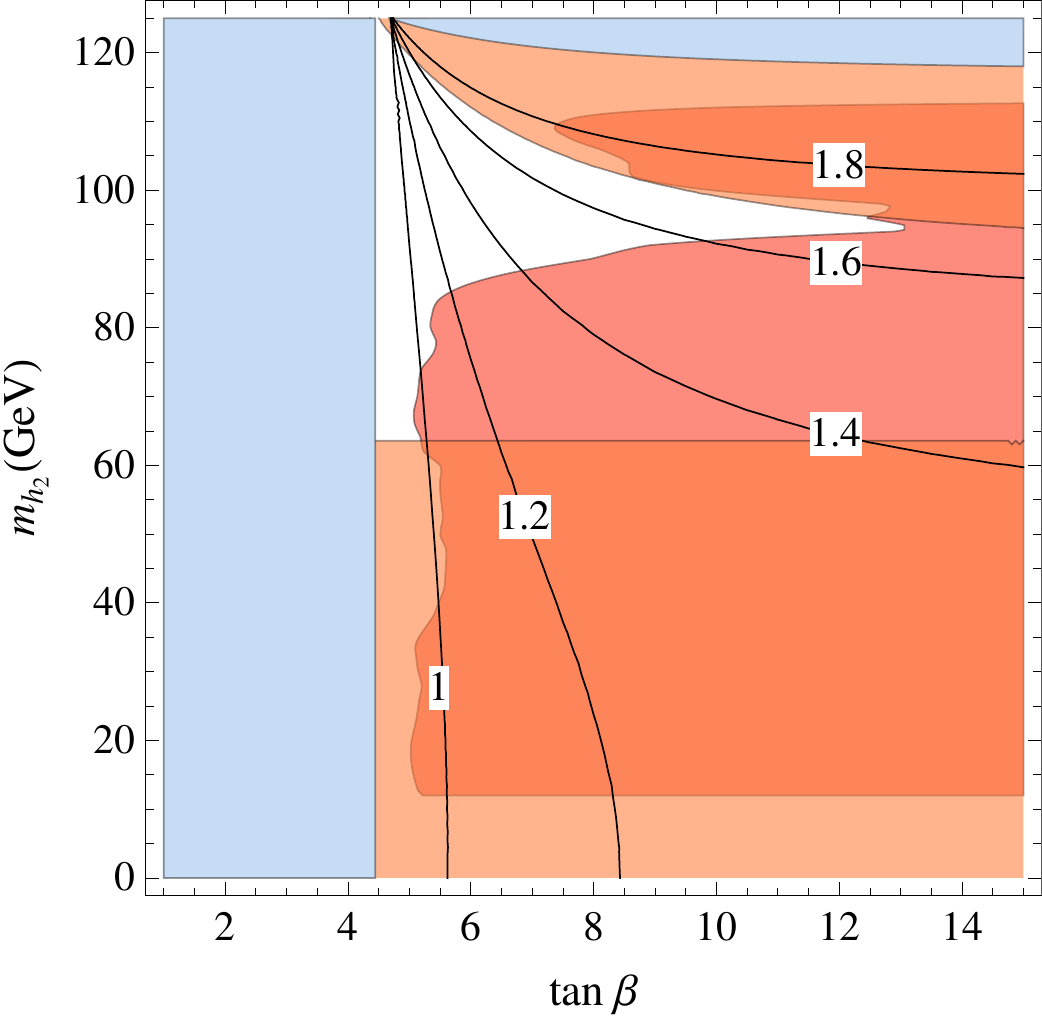}
\caption{\label{fig5} $H$-decoupled. Isolines of $g_{h^3}/g_{h^3}\big|_{\rm SM}$. Left: $\lambda=0.1$, $\Delta_t=85$ GeV and $v_S=v$. Right: $\lambda=0.8$, $\Delta_t=75$ GeV and $v_S=v$. Orange and blue regions as in Fig.~\ref{fig1}. The red region is excluded by LEP direct searches for $h_2\to b\bar b$.}
\end{center}
\end{figure}

How will it be possible to explore the regions of parameter space currently still allowed in this $h_2 < h_{\text{LHC}} (< h_3(= H))$ case in view of the reduced couplings of the lighter state? Unlike in the singlet-decoupled case, the improvement in the measurements of the signal strengths of $h_{\text{LHC}}$ is not going to play a major role. Based on the projected sensitivity of Table \ref{tab1}, the bound on the mixing angle will be reduced to $s^2_\gamma < 0.15$ at $95\%$ C.L. A significant  deviation from the case of the SM can occur in the cubic $h_{\text{LHC}}$-coupling, $g_{h^3_1}$, as shown in Fig.~\ref{fig5}.
The LHC14 in the high-luminosity regime is expected to get enough sensitivity to be able to see such deviations \cite{Dolan:2012rv,ATLAS-collaboration:2012iza,Goertz:2013kp}. 
At that point, on the other hand,  the searches for directly produced s-partners should have already given some clear indications on the relevance of the entire picture.

For completeness we recall from Ref. \cite{Barbieri:2013hxa} that the parameter space in the case $h_{\text{LHC}} < h_2 (< h_3(= H))$ is still largely unexplored at $\lambda = 0.7\div 1$. Most promising in this case are the direct searches of $h_2$ with gluon-fusion production cross-sections at LHC14 in the picobarn range and a large branching ratio, when allowed by phase space, into a pair of $h_{\text{LHC}}$. Furthermore here as well large deviations from the SM value can occur in the cubic $h_{\text{LHC}}$-coupling.

\section{Fully mixed case and the $\gamma \gamma$ signal}
\begin{figure}[t]
\begin{center}
\includegraphics[width=0.48\textwidth]{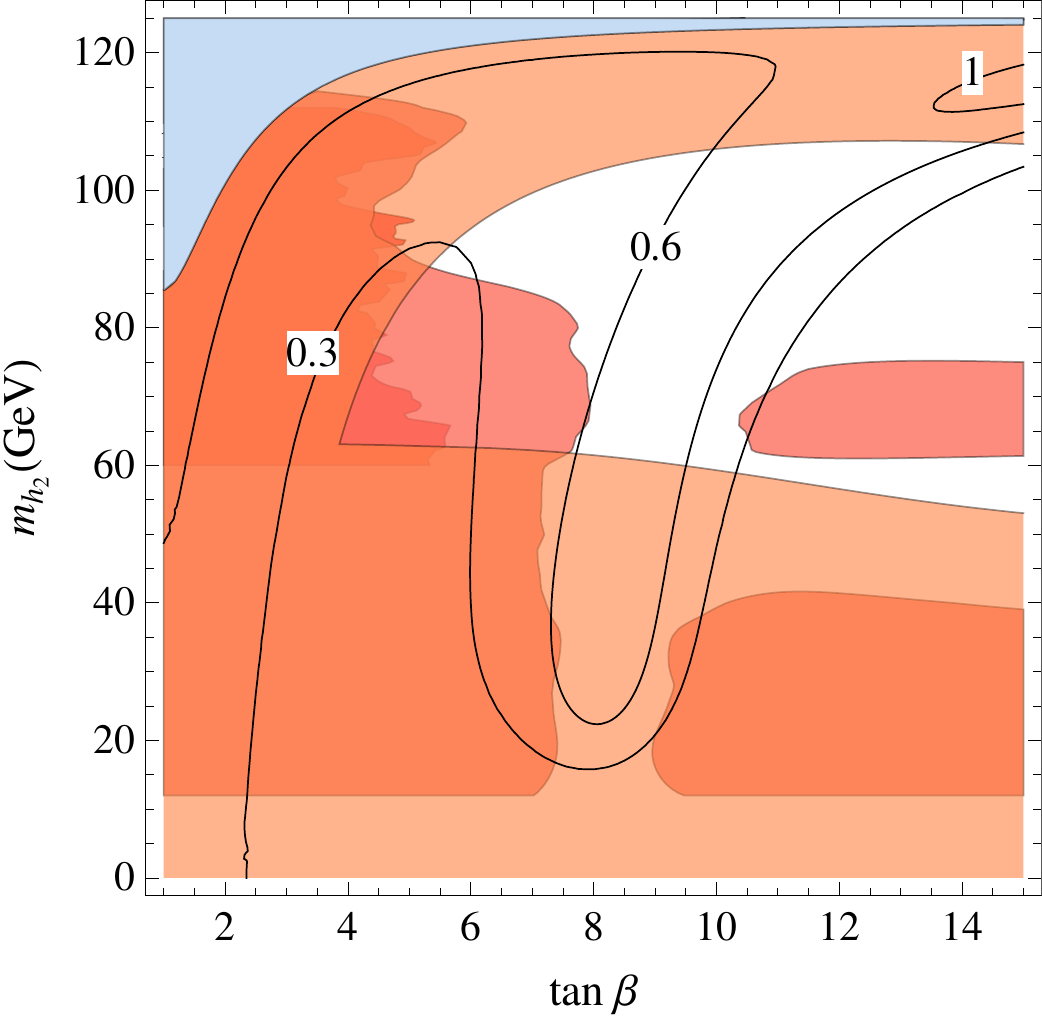}\hfill
\includegraphics[width=0.48\textwidth]{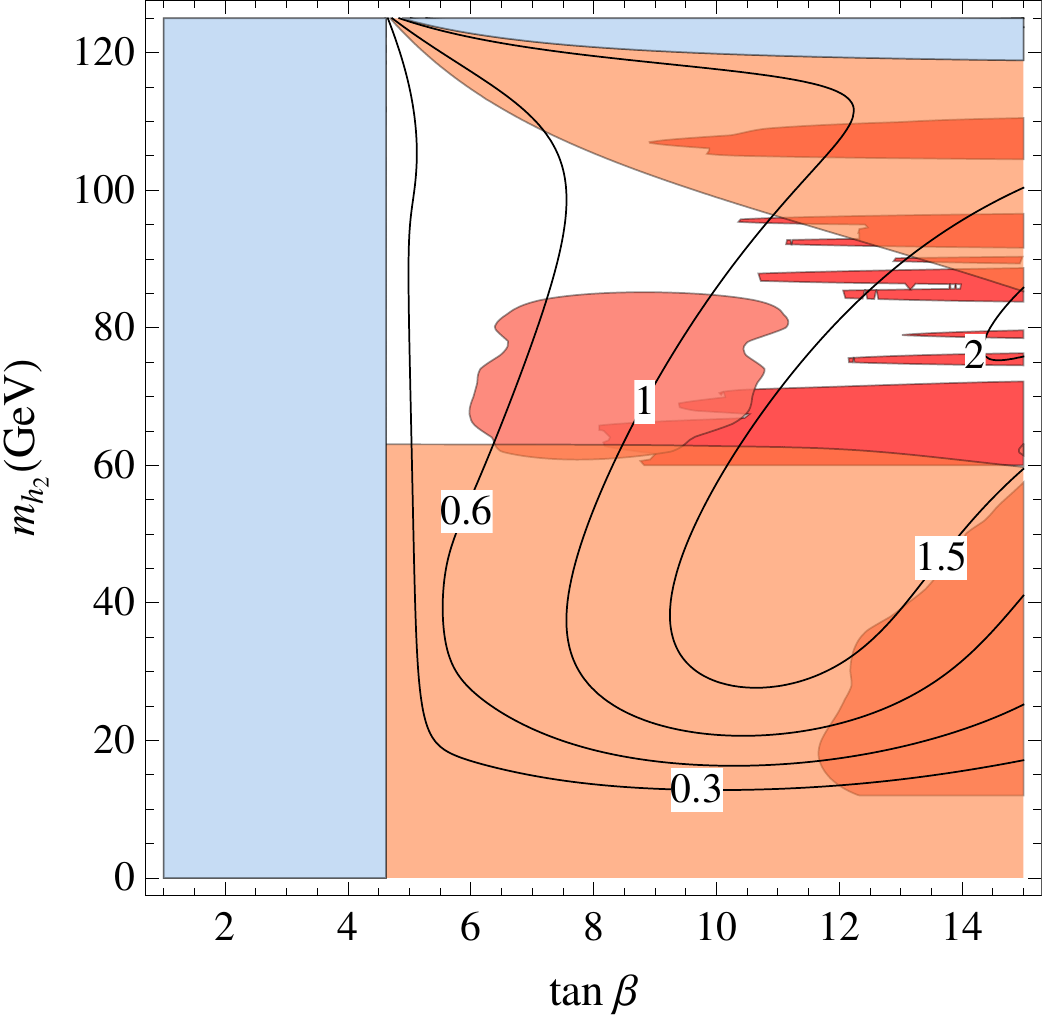}
\caption{\label{fig6} Fully mixed situation. Isolines of the signal strength of $h_2 \to \gamma \gamma$ normalized to the SM. We take  $m_{h_3}=500$ GeV, $s^2_\sigma=0.001$ and $v_s = v$. Left: $\lambda=0.1$, $\Delta_t=85$ GeV. Right: $\lambda=0.8$, $\Delta_t=75$ GeV. Orange and blue regions as in Fig.~\ref{fig1}. The red and dark red regions are excluded by LEP direct searches for $h_2\to b\bar b$ and $h_2 \to$ hadrons respectively.}
\end{center}
\end{figure}

The phenomenological exploration of the situation considered in the previous section could be significantly  influenced if the third state, i.e. the doublet $H$, were not fully decoupled. As an example we still consider the case of a state $h_2$ lighter than $h_{\text{LHC}}$, lowering $m_{h_3}$ to 500 GeV, to see if it could have an enhanced signal strength into $\gamma \gamma$.
Using Eqs.~\eqref{eq:sin:gamma:general}-\eqref{eq:sin:2alpha:general}, for fixed values of $\sigma$, $\lambda$ and $\Delta_t$, the two remaining angles $\alpha$ (or $\delta = \alpha - \beta + \pi/2$) and $\gamma$ are determined in any point of the $(\tan{\beta}, m_{h_2})$ plane and so are all the branching ratios of $h_2$ and of $h_{\text{LHC}}$. More precisely $\delta$ is fixed up to the sign of $s_\sigma c_\sigma s_\gamma$ (see first line of Eq.~\eqref{eq:sin:2alpha:general}), which is the only physical sign that enters the observables we are considering.

The corresponding situation is represented in Fig.~\ref{fig6}, for two choices of $\lambda$ and $\Delta_t$ (the choice $\lambda = 0.1$ was recently discussed in \cite{Badziak:2013bda}). The sign of $s_\sigma c_\sigma s_\gamma$ has been taken negative in order to suppress BR$(h_2\to b \bar b)$. This constrains $s_\sigma^2$ to be very small in order to leave a region still not excluded by the signal strengths of $h_{\text{LHC}}$, with $\delta$ small and negative. To get a signal strength for $h_2\rightarrow \gamma \gamma$ close to the SM one for the corresponding mass is possible for a small enough value of $s^2_\gamma$, while the dependence on $m_{h_3}$ is weak for values of $m_{h_3}$ greater than 500 GeV. Note that the suppression of the coupling of $h_2$ to $b$-quarks makes it necessary to consider the negative LEP searches for $h_2 \to$ hadrons \cite{Searches:2001aa}, which have been performed down to $m_{h_2} = 60$~GeV.

Looking at the similar problem when $h_2 > h_{\text{LHC}}$, we find it harder to get a signal strength close to the SM one, although this might be possible for a rather special choice of the parameters.\footnote{An increasing significance of the excess found by the CMS \cite{CMShint} at 136 GeV would motivate such special choice.} Our purpose here is more to show that in the fully mixed situation the role of the measured signal strengths of $h_{\text{LHC}}$, either current or foreseen, plays a crucial role.

\section{Electro-Weak Precision Tests}
\label{sec5}

One may ask if the electro-weak precision tests (EWPT) set some further constraint on the parameter space explored so far. We have directly checked that this is not the case in any of the different situations illustrated in the various figures. The reason is different in the singlet-decoupled and in the $H$-decoupled cases.

In the $H$-decoupled case the reduced couplings of $h_{\text{LHC}}$ to the weak bosons lead to well-known asymptotic formulae for the corrections to the $\hat{S}$ and $\hat{T}$ parameters \cite{Barbieri:2007bh}
\begin{equation}
\Delta \hat S =  + \frac{\alpha}{48\pi s_w^2}s^2_\gamma \log \frac{m_{h_2}^2}{m_{h_{\text{LHC}}}^2}    ,~~\Delta \hat T =      - \frac{3\alpha}{16\pi c_w^2}s^2_\gamma \log \frac{m_{h_2}^2}{m_{h_{\text{LHC}}}^2}
\end{equation}
valid for $m_{h_2}$ sufficiently heavier that $h_{\text{LHC}}$. The correlation of $s^2_\gamma$ with $m_{h_2}$ given in Eq. \eqref{sin2gamma} leads therefore to a rapid decoupling of these effects. 
The one loop effect on $\hat{S}$ and $\hat{T}$ becomes also vanishingly small as 
 $m_{h_2}$ and  $h_{\text{LHC}}$ get close to each other, since in the degenerate limit any mixing can be redefined away and only the standard doublet contributes as in the SM.

In the singlet-decoupled case the mixing between the two doublets can in principle lead to more important effects, which are however limited by the constraint on the mixing angle $\alpha$ or the closeness to zero of $\delta = \alpha - \beta +\pi/2$ already demanded by the measurements of the signal strengths of $h_{\text{LHC}}$.\footnote{Notice that in the fully mixed situation there may be relevant regions of the parameter space still allowed by the fit with a largish $\delta$ (see e.g. Fig. 1 of Ref. \cite{Barbieri:2013hxa}).  This could further constrain the small allowed regions, but the precise contributions to the EWPT depend on the value of the masses of the $CP$-odd scalars, which in the generic NMSSM are controlled by further parameters.}
Since in the $\delta = 0$ limit every extra effect on $\hat{S}$ and $\hat{T}$  vanishes, this explains why the EWPT do not impose further constraints on the parameter space that we have considered.

\section{The MSSM for comparison}
\label{sec6}

\begin{figure}[t]
\begin{center}
\includegraphics[width=0.48\textwidth]{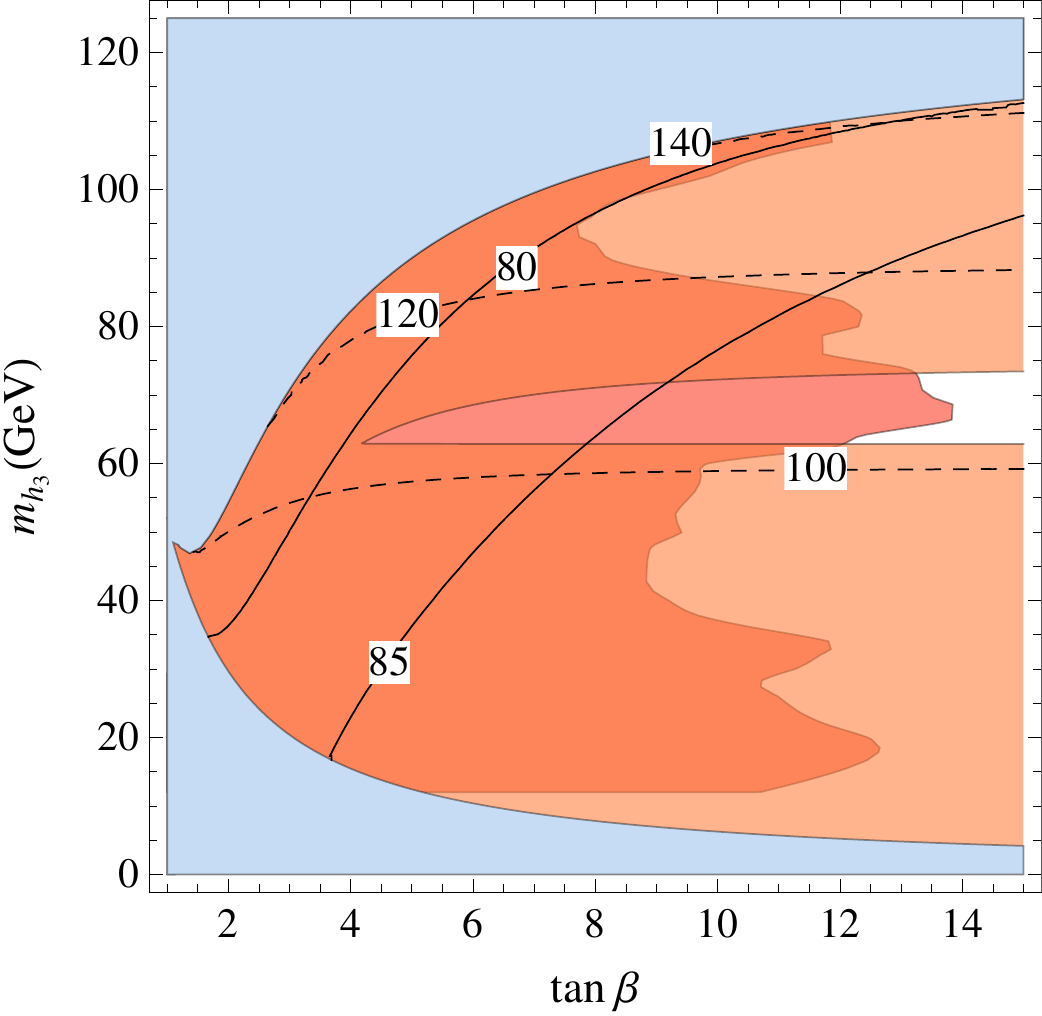}\hfill
\includegraphics[width=0.48\textwidth]{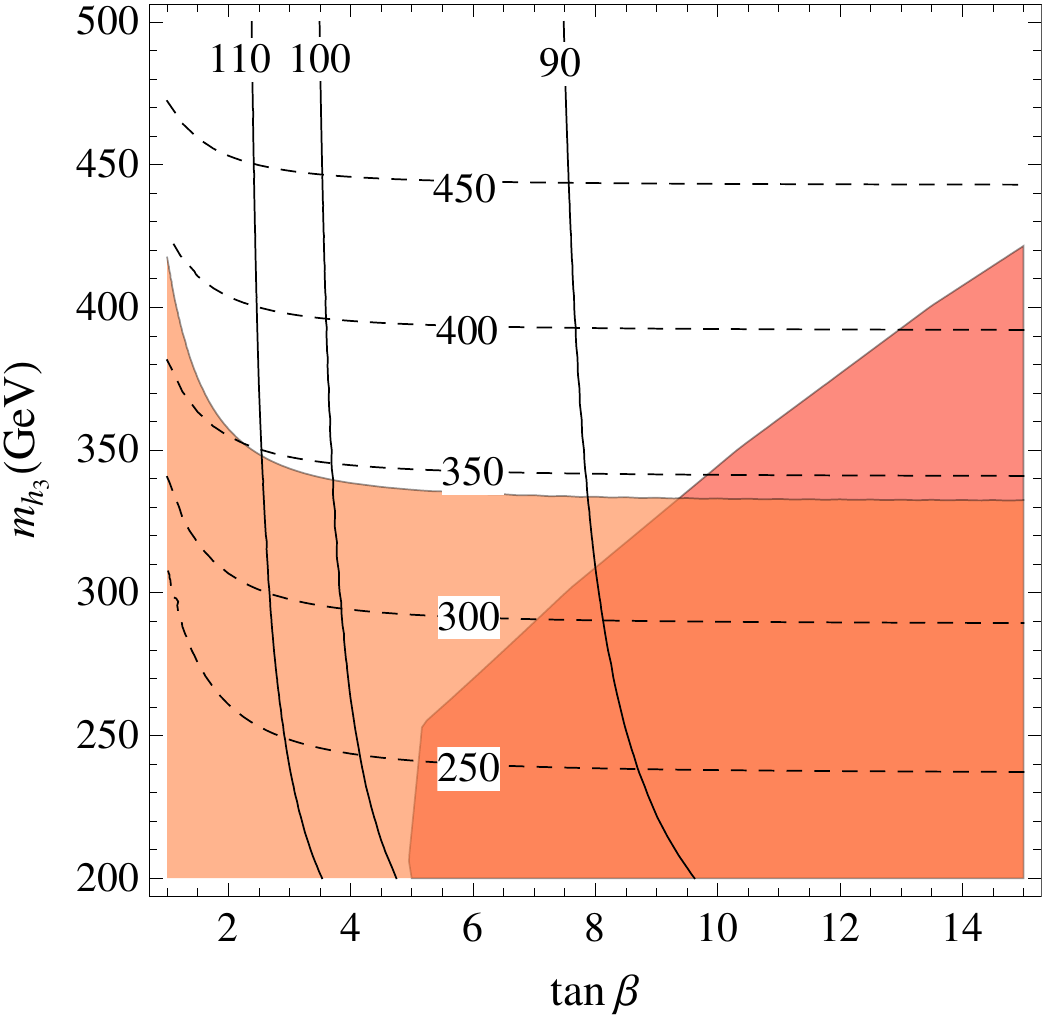}
\caption{\label{fig7} MSSM. Isolines of $\Delta_t$ (solid) and $m_{H^\pm}$ (dashed) at $(\mu A_t)/\langle m_{\tilde{t}}^2\rangle \ll 1$. Left: $h_{\rm LHC}>h_3$, red region is excluded by LEP direct searches for $h_3 \to b\bar b$. Right: $h_{\rm LHC}<h_3$, red region is excluded by CMS direct searches for $A, H\to \tau^+\tau^-$ \cite{CMS-PAS-HIG-12-050}. Orange and blue regions as in Fig.~\ref{fig1}.}
\end{center}
\end{figure}

As recalled in section \ref{sec3}, it is interesting to consider the MSSM, i.e. the $\lambda = 0$ limit
of the NMSSM in the singlet-decoupled case, using  as much as possible the
same language. The analogue of Fig. \ref{fig1} are shown in Fig. \ref{fig7}. From the point of view of the parameter space the main difference is that instead of $\lambda$ we use $\Delta_t$ as an effective parameter. As expected, both the left and right panel of Fig. \ref{fig7} make clear that a large value of $\Delta_t$ is needed to make the MSSM consistent with a 125 GeV Higgs boson.

At the same time, and even more than in the NMSSM case, the projection of the measurements of the signal strengths of 
$h_{\text{LHC}}$ is expected to scrutinize most of the parameter space. We have checked that this is indeed the case with the indirect  sensitivity to $m_{h_3}$ in the right panel of Fig. \ref{fig7}, which will be excluded up to about 1 TeV, as well as with the closure of the white region in the left side of the same Figure.
Notice that a similar exclusion will hold also for the CP-odd and charged Higgs bosons, whose masses are fixed in terms of the one of $h_3$.
A warning should be kept in mind, however, relevant to the case $h_3 < h_{\text{LHC}}$:  the one loop corrections to the mass matrix controlled by $(\mu A_t)/\langle m_{\tilde{t}}^2\rangle$ modify the left side of Fig. \ref{fig7} for $(\mu A_t)/\langle m_{\tilde{t}}^2\rangle \gtrsim 1$, changing in particular the currently and projected allowed regions.

\section{Summary and conclusions}
\label{sec7}

Given the current experimental informations, the Higgs sector of the NMSSM appears to allow a minimally fine-tuned description of electro-weak symmetry breaking, at least in the context of supersymmetric extensions of the SM. Motivated by this fact and complementing Ref. \cite{Barbieri:2013hxa}, we have outlined a possible overall strategy to search for signs of the $CP$-even states by suggesting a relatively simple analytic description of  four different  situations:
\begin{itemize}
\item Singlet-decoupled, $h_3  < h_{\rm LHC} < h_2 (= S)$
\item Singlet-decoupled, $h_{\rm LHC} < h_3  < h_2 (= S)$
\item $H$-decoupled, $h_2  < h_{\rm LHC} < h_3 (= H)$
\item $H$-decoupled, $h_{\rm LHC} < h_2  < h_3 (= H)$
\end{itemize}
To make this possible at all we have made some simplifying assumptions on the parameter space, which are motivated by naturalness requirements and have been in any case specified whenever needed.
In our view the advantages of having an overall coherent analytic picture justify the introduction of these assumptions.

Not surprisingly,  a clear difference emerges between the singlet-decoupled and the $H$-decoupled cases: the influence on the signal strengths of  $h_{\text{LHC}}$ of the mixing with a doublet or with a singlet makes the relative effects visible at different levels. A quantitative estimate of the sensitivity of the foreseen measurements at LHC14 with 300 $\text{fb}^{-1}$ makes it likely that the singlet-decoupled case will be thoroughly explored, while the singlet-mixing effects could remain hidden.
We also found that, in the MSSM with $(\mu A_t)/\langle m_{\tilde{t}}^2\rangle \lesssim 1$, the absence of deviations in the $h_{\text{LHC}}$ signal strengths would push the mass of the other Higgs bosons up to a TeV. Needless to say, in any case the direct searches will be essential with a variety of possibilities discussed in the literature. As an example we have underlined the significance of $h_2\rightarrow   h_{\text{LHC}} h_{\text{LHC}}$ in the $h_{\rm LHC} < h_2  < h_3 (= H)$ case. It is also interesting that, in the $H$-decoupled case, large deviations from the SM value are possible in the triple Higgs coupling $g^3_{h_{\rm LHC}}$, contrary to the $S$-decoupled and MSSM cases.
More in general  it is useful to observe that  the framework outlined in this work makes possible to describe the impact of the various direct searches in a systematic way, together with the indirect ones in the $h_{\text{LHC}}$ couplings. 
Finally, in case of a positive signal, direct or indirect, it may be important to try to interpret it in a fully mixed scheme, involving all the three $CP$-even states. To this end the analytic relations of the mixing angles to the physical masses given in Eqs.~\eqref{eq:sin:gamma:general}-\eqref{eq:sin:2alpha:general} offer a useful tool, as illustrated in the examples of a $\gamma\gamma$ signal of Fig.~\ref{fig6}.

It will be interesting to follow the progression of the searches of the Higgs system of the
NMSSM, directly or indirectly through the more precise measurements of the  properties of the state already found at the LHC.

\subsubsection*{Acknowledgments}
We would like to thank Pietro Slavich for useful discussions.
This work is supported in part by the European Programme ``Unification in the LHC Era",  contract PITN-GA-2009-237920 (UNI\-LHC), MIUR under the contract 2010YJ2NYW-010, the ESF grants 8943, MJD140 and MTT8, by the recurrent financing SF0690030s09 project and by the European Union through the European Regional Development Fund. 
We would like to thank the Galileo Galilei Institute in Florence for hospitality during the completion of this work.

\bibliographystyle{My}
\small
\bibliography{Higgses_arxiv_v2}

%\begin{thebibliography}{99}     
%
%\end{thebibliography}
\end{document}